\documentstyle[preprint,tighten,aps,amstex,graphicx,times,floats]{revtex}

\begin{document}

\thispagestyle{empty}

% macros for marking changes
\marginparwidth 1.cm
\setlength{\hoffset}{-1cm}

\renewcommand{\abstractname}{Abstract}
\renewcommand{\figurename}{Figure}
\renewcommand{\tablename}{Table}
\renewcommand{\refname}{Bibliography}

% additional commands 
\newcommand{\eg}{{\it e.g.}\;}
\newcommand{\ie}{{\it i.e.}\;}
\newcommand{\etal}{{\it et al.}\;}
\newcommand{\ibid}{{\it ibid.}\;}

\newcommand{\mx}{M_{\rm SUSY}}
\newcommand{\pt}{p_{\rm T}}
\newcommand{\et}{E_{\rm T}}
\newcommand{\del}{\varepsilon}
\newcommand{\sla}[1]{/\!\!\!#1}
\newcommand{\fb}{\;{\rm fb}}
\newcommand{\gev}{\;{\rm GeV}}
\newcommand{\tev}{\;{\rm TeV}}
\newcommand{\abi}{\;{\rm ab}^{-1}}
\newcommand{\fbi}{\;{\rm fb}^{-1}}

\preprint{
\font\fortssbx=cmssbx10 scaled \magstep2
\hbox to \hsize{
\hfill\vtop{\hbox{\bf CERN-TH/2003-068}
            \hbox{\today}                    } }
}

\title{ 
Higgs--Boson Production Induced by Bottom Quarks 
} 

\author{
Eduard Boos$^1$ and Tilman Plehn$^{2}$
} 

\address{ 
$^1$ Skobeltsyn Institute of Nuclear Physics, Moscow State University,
     Moscow, Russia \\
$^2$ Theory Division, CERN, Geneva, Switzerland} 

\maketitle 

\begin{abstract}
Bottom quark--induced processes are responsible for a large fraction of
the LHC discovery potential, in particular for supersymmetric Higgs
bosons. Recently, the discrepancy between exclusive and inclusive
Higgs boson production rates has been linked to the choice of an
appropriate bottom factorization scale. We investigate the process
kinematics at hadron colliders and show that it leads to a
considerable decrease in the bottom factorization scale. This effect
is the missing piece needed to understand the corresponding higher
order results. Our results hold generally for charged and for neutral
Higgs boson production at the LHC as well as at the Tevatron. The
situation is different for single top quark production, where we find
no sizeable suppression of the factorization scale. Turning the argument
around, we can specify how large the collinear logarithms are, which
can be resummed using the bottom parton picture.
\end{abstract} 

\vspace{0.2in}

%%%%%%%%%%%%%%%%%%%%%%%%%%%%%%%  MAIN TEXT  %%%%%%%%%%%%%%%%%%%%%%%%%%%%

\section{Higgs Bosons at the LHC}

The combined LEP precision measurements~\cite{lep_precision} suggest
the existence of a light Higgs boson. In the case of a single Standard
Model Higgs boson the LHC promises multiple coverage for any Higgs
boson mass, which will enable us to measure its different decay modes
and extract the couplings~\cite{lhc_sm}. For a supersymmetric Higgs
sector this coverage has to rely on fewer Higgs boson decay
channels~\cite{lhc_sm,lhc_susy}. This is a direct consequence of the
structure of the Higgs sector: while the Minimal Supersymmetric
Standard Model (MSSM) predicts a light Higgs boson, it also predicts
an enhancement of the coupling to down-type fermions, at the expense
of the branching fractions to gauge bosons. This enhancement is an
outcome from the two Higgs doublet structure in the MSSM: one doublet
is needed to give mass to up-type, the other to down-type
fermions. The vacuum expectation values of the two doublets are
different, parameterized by $\tan \beta = v_2/v_1$. In addition to a
light scalar Higgs boson, the two Higgs doublet model includes a heavy
scalar, a pseudoscalar, and a charged Higgs boson.  None of these
additional particles have a mass bounded from above, apart from
triviality or unitarity bounds.

Of course, observables linked to properties of a light Higgs boson can
serve as a probe if a new scalar particle is indeed consistent with
the Standard Model Higgs boson~\cite{tesla,sven}. There is, however,
only one way to conclusively tell the supersymmetric Higgs sector from
its Standard Model counterpart: to discover the additional heavy Higgs
bosons and determine their properties.\medskip

At the LHC, the possible enhancement of down-type fermion Yukawa
couplings by powers of $\tan\beta$ can render the search for a heavy
scalar and pseudoscalar Higgs boson promising. For small values of
$\tan\beta$ the Yukawa coupling of the charged Higgs boson is governed
by the top quark mass $m_t/\tan\beta$, whereas for larger values of
$\tan\beta$ the bottom Yukawa coupling $m_b \tan\beta$
dominates. While the chances of finding a heavy Higgs boson with a
small value of $\tan\beta$ at the LHC are rather slim, the discovery
of all heavy Higgs scalars in the large $\tan\beta \gtrsim 10$ regime
is likely. This reflects the fact that the reach of the LHC 
for charged and neutral Higgs bosons is to a large degree owed to
scattering processes which involve incoming bottom quarks. The
completely exclusive processes are
\begin{equation}
gg \to \bar{b} t H^- \qquad \qquad
gg \to \bar{b} b \Phi,
\label{eq:sig_excl}
\end{equation}
where $\Phi=h^0,H^0,A^0$ denotes any neutral scalar Higgs boson.  From
an experimental point of view, the identification of these final state
bottom jets is tedious, because the dominant contribution comes from
phase space configurations where the incoming gluons split into two
collinear bottom quarks. The bottom quark rapidity distribution peaks
at rapidities around two and the transverse momentum distributions
around the bottom quark mass. After adding in the efficiency for a
bottom tag this becomes a heavy price to pay in the
analysis. Therefore one usually prefers to look for more or less
inclusive channels
\begin{equation}
gb \to t H^- \qquad \qquad
gb \to b \Phi \qquad \qquad
b\bar{b} \to \Phi.
\label{eq:sig_incl}
\end{equation}
At this point we do not explicitly discuss the bottom--induced
inclusive channels $b\bar{b} \to W^+H^-$~\cite{charged_bb} and $bg \to
tW^-$~\cite{zhu}, which are both known to next-to-leading order QCD,
because their impact on the discovery potential of the LHC is not
drastic. We emphasize, however, that our argument will hold for them
the same way we apply it to the processes in
eq.(\ref{eq:sig_incl}).\medskip

All bottom--inclusive channels suffer from an additional uncertainty:
the choice of the factorization scale of the bottom parton. In
contrast to the gluon density, the dependence of the bottom parton
density on the factorization scale is large even for scales above
${\cal O}(100\gev)$. Recently, it has been observed in higher order
calculations~\cite{charged_bb,zhu,mine,scott2,neut_bb_nnlo}, that
varying the bottom factorization scale around a smaller central value
yields a more stable perturbative behavior. The same choice of scales
can resolve the discrepancy between the total inclusive and exclusive
rate, which is most prominent for the production channel $b\bar{b} \to
\Phi$. However, choosing scales according to perturbative behavior is
difficult, because most processes include inherent cancellations
between different contributions, and it is hard to define which
contributions actually have to be stable. Instead of this somewhat
soft argument, we will derive an appropriate bottom factorization
scale from the kinematics of the exclusive production process.  In the
following two sections we first investigate charged Higgs
boson production, because it involves only one incoming bottom quark
and two heavy central decay products. In Section~\ref{sec:others} we
will then generalize our result to neutral Higgs boson production and
compare it to single top quark production.\medskip

\underline{Conventions:} Throughout this paper we show consistent
leading order cross section predictions, including the respective one
loop coupling constant, running heavy quark masses, and CTEQ parton
densities. Unless stated otherwise, we assume $\tan\beta=30$ for all
MSSM processes. The exclusive cross sections are quoted with a massive
$4.6\gev$ bottom quark in the matrix element and the phase space,
while the bottom Yukawa coupling is set to the running bottom quark
mass.

\section{Bottom Parton Scattering at the LHC} 
\label{sec:bottom_parton}

Heavy flavor--induced search channels for supersymmetric Higgs bosons
have been explored for many years~\cite{bottom_parton}. To begin with
the charged Higgs boson, three search modes have been investigated:
(1) charged Higgs bosons can be pair produced in a Drell--Yan type
process, mediated by a weak interaction
vertex~\cite{charged_dy}. Moreover, they can be pair produced at tree
level in bottom quark scattering~\cite{charged_bb} or through a one
loop amplitude in gluon fusion~\cite{charged_gg}. (2) One charged
Higgs boson can be produced together with a $W$ boson via scattering
of two bottom quarks or in gluon fusion~\cite{higgs_w_asso}. (3) The
charged Higgs boson can be produced in association with a top quark,
which seems to be the most promising search
channel~\cite{zhu,mine,charged_theo,charged_nlo}.  The charged Higgs
boson can be detected either decaying to a top and a bottom
quark~\cite{charged_top} or decaying to a tau lepton and a
neutrino~\cite{charged_tau}. Both LHC collaborations have tried to
reproduce these phenomenological analyses, most successfully in the
case of the decay to tau
leptons~\cite{charged_atlas,charged_cms}. However, these analyses have
to be taken with a grain of salt, if they rely on the standard Monte
Carlo generators, because some do not include a running bottom Yukawa
coupling. From the next-to-leading order QCD
calculations~\cite{zhu,mine} we know that using the bottom pole mass
as the Yukawa coupling severely overestimates the rates, as one would
expect from what we know about Higgs boson decays to bottom
quarks\footnote{We can compare these next-to-leading SUSY-QCD
calculations~\cite{zhu,mine} to the usual heavy parton subtraction
schemes which add the exclusive and the inclusive
channels~\cite{bottom_parton}: for large final state masses, the
higher order calculation of the inclusive channel is just the
perturbatively consistent extension of the latter.}.\smallskip

The case of neutral scalar Higgs bosons in the MSSM has been discussed
in similar detail. In the Standard Model the inclusive production
process $b\bar{b} \to \Phi$ is a small correction to the inclusive
gluon fusion channel~\cite{neut_gg}. In a supersymmetry context, and
in particular for large $\tan\beta$, the bottom quark--induced process
dominates gluon
fusion~\cite{lhc_sm,neut_bb_nnlo,neut_bb_excl,neut_bb_incl}. The
typical Higgs decays are the same as for a light Standard Model Higgs
boson, except for heavy MSSM scalars, where decays to muon or tau
pairs are most promising~\cite{neut_heavy}. Additional problems occur
in the so-called intense coupling regime, \ie the region with an
intermediate pseudoscalar Higgs boson mass and strong mixing
effects~\cite{lhc_susy,intense}. The two scalar Higgs bosons can be
detected and possibly be resolved in weak boson fusion with a
subsequent decay to $\tau$ pairs~\cite{lhc_susy}; unfortunately, for
large values of $\tan\beta$ the mass splitting becomes too small
to distinguish the two mass peaks. In that region, the most promising
way to search for and separate the heavy Higgs states is the production
process involving $b$ jets with the Higgs bosons decaying to
muons~\cite{intense2}. To distinguish the gluon fusion process from
the bottom quark--induced production, one can require observation of only one
final state bottom quark, \ie using the partly inclusive channel $gb \to
b \Phi$~\cite{scott2,scott1}.
\bigskip

Motivated by this vast number of analyses in the MSSM Higgs sector we
turn to the simplest process available: charged Higgs boson production
in association with a top quark involves only one incoming bottom quark
and is an appropriate starting point to understand the issue of bottom
partons at the LHC. The features of the exclusive and the inclusive
charged Higgs boson production process
\begin{equation}
gg \to \bar{b} t H^- \qquad \qquad \qquad
gb \to t H^-
\end{equation}
have been investigated in detail in Ref.~\cite{mine}. Let us briefly
review the most important observations:\smallskip

For reasons described above, the searches for charged Higgs bosons
(decaying to tau leptons) at the LHC usually do not require the
observation of a final state bottom quark. The exclusive process $gg
\to \bar{b} t H^-$ is then dominated by collinear splitting of one of
the gluons into a bottom quark pair:
\begin{equation}
\frac{d \sigma}{d p_{T,b}} \Bigg|_{\rm asympt} \propto 
\frac{p_{T,b}}{p_{T,b}^2+m_b^2}  \qquad \qquad \qquad \qquad \qquad
\sigma \Big|_{\rm asympt} \propto 
\log \left[ \left( \frac{p_{T,b}^{\rm max}}{m_b} \right)^2 
                         + 1 \right]
\label{eq:excl_asymp}
\end{equation}
These logarithms in the total cross section can be resummed, which is
precisely the definition of bottom partons~\cite{bottom_parton}. The
factorization scale of the bottom partons is defined as the maximum
transverse bottom quark momentum up to which the asymptotic form of
the cross section is assumed to hold, and up to which the logarithms
$\log p_{T,b}/m_b$ are then resummed. This means that in
eq.(\ref{eq:excl_asymp}) one can identify $p_{T,b}^{\rm max}$ with
$\mu_{F,b}$. If we assume $p_{T,b}^{\rm max} \gg m_b$, which as we
will see is true for all processes we consider, the numerical value
for the bottom quark mass will have no impact on our argument --- we
could neglect it altogether~\cite{mine,scott2,bottom_parton}. The
factorization scale is {\sl per se} an artificial parameter which has
to vanish after including all orders of perturbation theory. In the
case of neutral Higgs boson production $b\bar{b}\to \Phi$ this has
recently been demonstrated, including the NNLO QCD
corrections~\cite{neut_bb_nnlo}. We are lucky in the case of bottom
quarks: the comparison between the actual and the asymptotic form of
the {\sl exclusive} cross section allows us to estimate $p_{T,b}^{\rm
max}$ and therefore the bottom parton factorization scale for the {\sl
inclusive} process. This also tells us how large the `large
logarithms', which are resummed using bottom parton densities,
actually are.\medskip

There are two caveats to be kept in mind, though. First, the inclusive
process $gb \to t H^-$ assumes that the entire cross section comes
from a phase space region in which we can neglect the transverse
momentum of the bottom jet appearing in the exclusive process $gg \to
\bar{b} t H^-$. This is probably a good approximation, in particular
after including detector effects, but one has to be aware of
it. Second, naive dimensional analysis suggests $\mu_{F,b} = m_t +
m_H$, which is often used. This does not have to be correct. The only
thing dimensional analysis tells us is $\mu_{F,b} \propto (m_t +
m_H)$, as long as the production process is dominated by the threshold
region. It has been shown that the proportionality factor does not at
all have to be unity and that a wrong choice of scale leads to a
systematic overestimate of the cross section, as it is obvious from
eq.(\ref{eq:excl_asymp})~\cite{mine,scott2,dieter}.

\section{Asymptotic Behavior in the Bottom Quark Virtuality} 
\label{sec:plateau_q}

\begin{figure}[t] 
\begin{center}
\includegraphics[width=14.0cm]{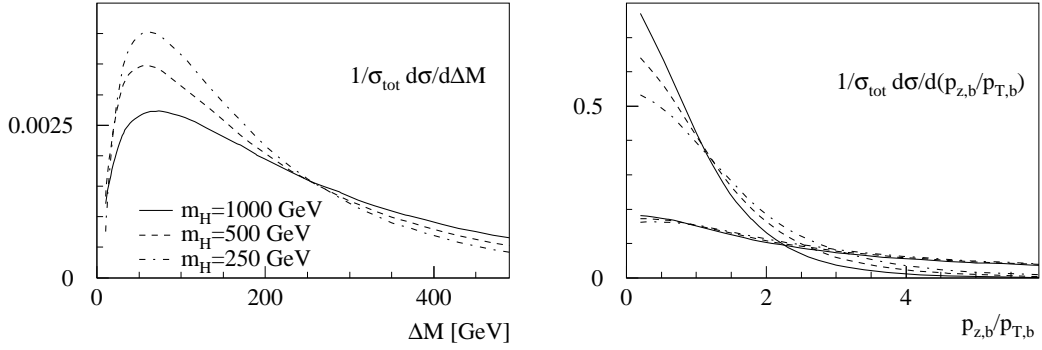}
\end{center}
\vspace*{0mm}
\caption[]{\label{fig:para_bth} Normalized distributions for the
hadronic charged Higgs boson production $gg \to \bar{b}tH^-$. Left:
difference between the invariant mass of the $tH^-$ system and its
threshold mass $M=m_t+m_H$. Right: ratio of the longitudinal and
transverse momentum of the bottom jet $\rho = p_{z,b}/p_{T,b}$ in the
parton center-of-mass system.  The steeper set of curves is after a
cut $Q_b>M/5$.}
\end{figure}

To make our argument as transparent as possible, we will investigate
the relation between the bottom factorization scale and the threshold
mass in two steps. The threshold mass we will refer to as $M$; for the
charged Higgs boson production process this means $M = m_t+m_H$.  We
can generally rewrite the exclusive and the inclusive production
processes including one bottom parton as
\begin{equation}
gg \to \bar{b} X_M \qquad \qquad
bg \to X_M,
\label{eq:proc_gen}
\end{equation}
where $X_M$ denotes the heavy final state particles. For a typical
gluon--induced LHC production process close to threshold we expect the
invariant mass of the heavy system and the threshold mass $M$ to be
similar. In the case of $gg \to \bar{b}tH^-$ we investigate how close
to threshold the production takes place in
Fig.~\ref{fig:para_bth}. The parameter $\Delta M$ is defined as the
difference between the invariant mass of the heavy $t H^-$ system and
the threshold mass $M=m_t+m_H$. We indeed see that the distributions
peak at small values $\Delta M/M \lesssim 1/8$, even though there are
sizeable tails towards larger invariant masses.\bigskip

In the first step of our argument, we investigate the maximum
value for the intermediate bottom quark virtuality $Q_b^{\rm max}$, up
to which the asymptotic form of the exclusive cross section holds
\begin{equation}
Q_b^{\rm max} = C_Q \; M.
\end{equation}

In Section~\ref{sec:plateau_pt}, we then estimate how far the
asymptotic form in terms of the bottom transverse momentum is
valid. From Ref.~\cite{mine} we expect a relation $p_{T,b}^{\rm max}
\sim M/6$. If we assume that this reduction will consist of the
suppression from the asymptotic behavior in the virtuality and an
additional suppression when we move to the asymptotic behavior in the
transverse momentum, we obtain
\begin{equation}
\mu_{F,b} \equiv p_{T,b}^{\rm max} = C_p \; Q_b^{\rm max}  
= C_p \; C_Q \; M
\end{equation}\bigskip 

To understand the asymptotic behavior of the hadronic cross section
$\sigma(pp\hskip-7pt\hbox{$^{^{^{(\!-\!) }}}$} \to \bar{b}X_M)$ as a
function of the bottom quark virtuality, we rewrite the integration
over the phase space and the parton momentum
fractions~\cite{scott2}. As long as we are interested in the behavior
of the cross section for large values of the rapidity we can safely
neglect the bottom quark mass:
\begin{equation}
\sigma =    \frac{1}{16 \pi} \frac{1}{S}
            \int_0^{S-M^2} dQ_b^2 
            \int_{Q_b^2+M^2}^S ds 
            \int_{\frac{1}{2} \log \frac{s}{S}}^{-\frac{1}{2} \log \frac{s}{S}} dy \;
            {\cal L}_{gg} \;
            \frac{1}{s^2} \;
            \overline{|{\cal M}|^2} 
\label{eq:cxn1}
\end{equation}

\begin{figure}[t] 
\begin{center}
\includegraphics[width=8.0cm]{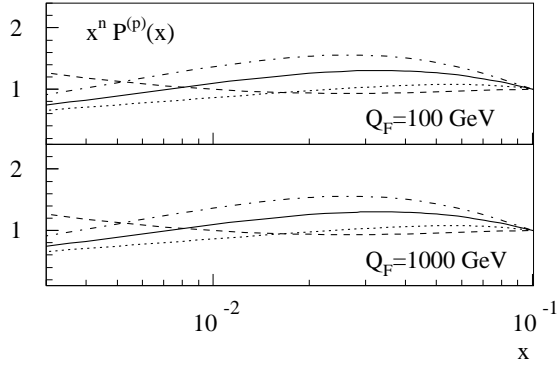}
\end{center}
\vspace*{0mm}
\caption[]{\label{fig:lumis} Ratio of parton distributions in the
proton and the function $x^{-n}$ for two different values of the
factorization scale. All curves are normalized to their values at
$x=0.1$. The different lines correspond to the gluon (solid, $n=2$),
down-quark (dashed, $n=1.1$), anti-up-quark (dotted, $n=1.7$), and
bottom quark(dash--dotted, $n=2$) content. We use CTEQ6L parton
densities~\cite{cteq6}.}
\end{figure}

\begin{figure}[t] 
\begin{center}
\includegraphics[width=7.0cm]{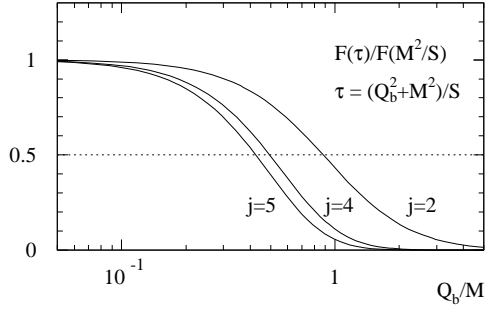}
\end{center}
\vspace*{0mm}
\caption[]{\label{fig:func} The normalized function $F(\tau)$, defined
in eq.(\ref{eq:cxn3}). The hadronic center-of-mass energy is set to
$\sqrt{S}=14\tev$ and the threshold mass to $675\gev$, corresponding
to a $500\gev$ charged Higgs boson. We display the behavior of the
plateau in $Q_b$ for different values of $j$, which arise from the
$x_1 x_2$ behavior of the partonic cross section.}
\end{figure}

Here $Q_b$ is the intermediate bottom quark virtuality, $y$ the
rapidity, and $\sqrt{S}$ and $\sqrt{s}$ are the hadronic and partonic
collider energies. The factor $1/s^2$ in the integrand is obvious from
the difference in mass units between the matrix element and the
partonic differential cross section $d\hat{\sigma}/dQ_b^2$, as it
originally appears in the integral. The parton densities are denoted
as ${\cal L}=P_{i/p}(x_1) P_{j/p}(x_2)$. At this point we make a few
simplifying approximations: since we want to show that the asymptotic
behavior in the virtuality is a process-independent phase space
effect, we neglect all structures in the matrix element, except for
the asymptotic behavior in the virtuality. The asymptotic form of the
differential hadronic cross section $d\sigma/dQ_b \propto 1/Q_b$
translates into $\overline{|{\cal M}|^2} = S^2 \sigma_0/Q_b^2$. The
factor $S^2$ we introduce to absorb units of energy, it could as well
be $M^4$. For reasons which will be obvious later, it could not be a
partonic variable, since we need to keep track of the powers of the
parton momentum fraction. Furthermore, we assume that the steep gluon
densities balance each other for the production of heavy states
$x_1=x_2=\sqrt{x}$ for $x_1,x_2 \ll 1$. The approximate hadronic cross
section now reads
\begin{alignat}{7}
\sigma  &=  - \frac{2\sigma_0}{16 \pi} \,
              \int_0^{S-M^2} \frac{dQ_b}{Q_b} 
              \int_{\frac{Q_b^2+M^2}{S}}^1 \frac{dx}{x^2} \;
              {\cal L}_{gg}(x) \;
              \log x \notag \\
        &=  \frac{2\sigma_0 {\cal L}_0}{16 \pi} \,
            \int_0^{S-M^2} \frac{dQ_b}{Q_b} \; F(\tau(Q_b^2))
\label{eq:cxn2}
\end{alignat}
using 
\begin{equation}
F(\tau)  = -\int_\tau^1 \frac{dx}{x^2} \;
           \frac{1}{x^{j-2}} \; \log x 
         = \frac{1}{(1-j)^2} \; \left[ 
           1 - \tau^{1-j} + (1-j) \, \tau^{1-j} \, \log \tau \right]
         \sim \; Q_b \frac{d\sigma}{dQ_b},
\label{eq:cxn3}
\end{equation}
with $x=s/S$ and $\tau=(Q_b^2+M^2)/S$. The function $F(\tau)$ is as a
correction to the asymptotic behavior of the virtuality distribution
$d\sigma/dQ_b \propto 1/Q_b$. In an intermediate step we have
approximated the incoming parton luminosity by a simple power
suppression ${\cal L} = {\cal L}_0/x^{j-2}$. As a general
parameterization of the parton densities, this is certainly not a good
idea. However, looking at the production of heavy particles at the LHC
we probe momentum fractions between $10^{-1}$ and few times
$10^{-3}$. In Fig.~\ref{fig:lumis} we show different parton
distributions multiplied by $x^n$ for two values of the factorization
scale. The ratio is normalized to its value at $x=0.1$. Looking at the
different values of $n$ we see that our argument will at this point
become dependent on the parton the incoming bottom parton sees on the
other side. On the other hand, we also see that for $gg$ and $gb$
initial states the approximation ${\cal L} \propto 1/(x_1 x_2)^2$
works very well. From Fig.~\ref{fig:lumis} we obtain $j=4$ for gluon
or bottom initial states in the definition of $F(\tau)$. In
Fig.~\ref{fig:func} we show the behavior of $F(\tau)$ as a function of
the bottom quark virtuality, the way it looks for a $500\gev$ charged
Higgs boson at the LHC. The case of $j=2$ corresponds to constant
parton densities ${\cal L} \equiv {\cal L}_0$. \bigskip

\begin{figure}[t] 
\begin{center}
\includegraphics[width=14.0cm]{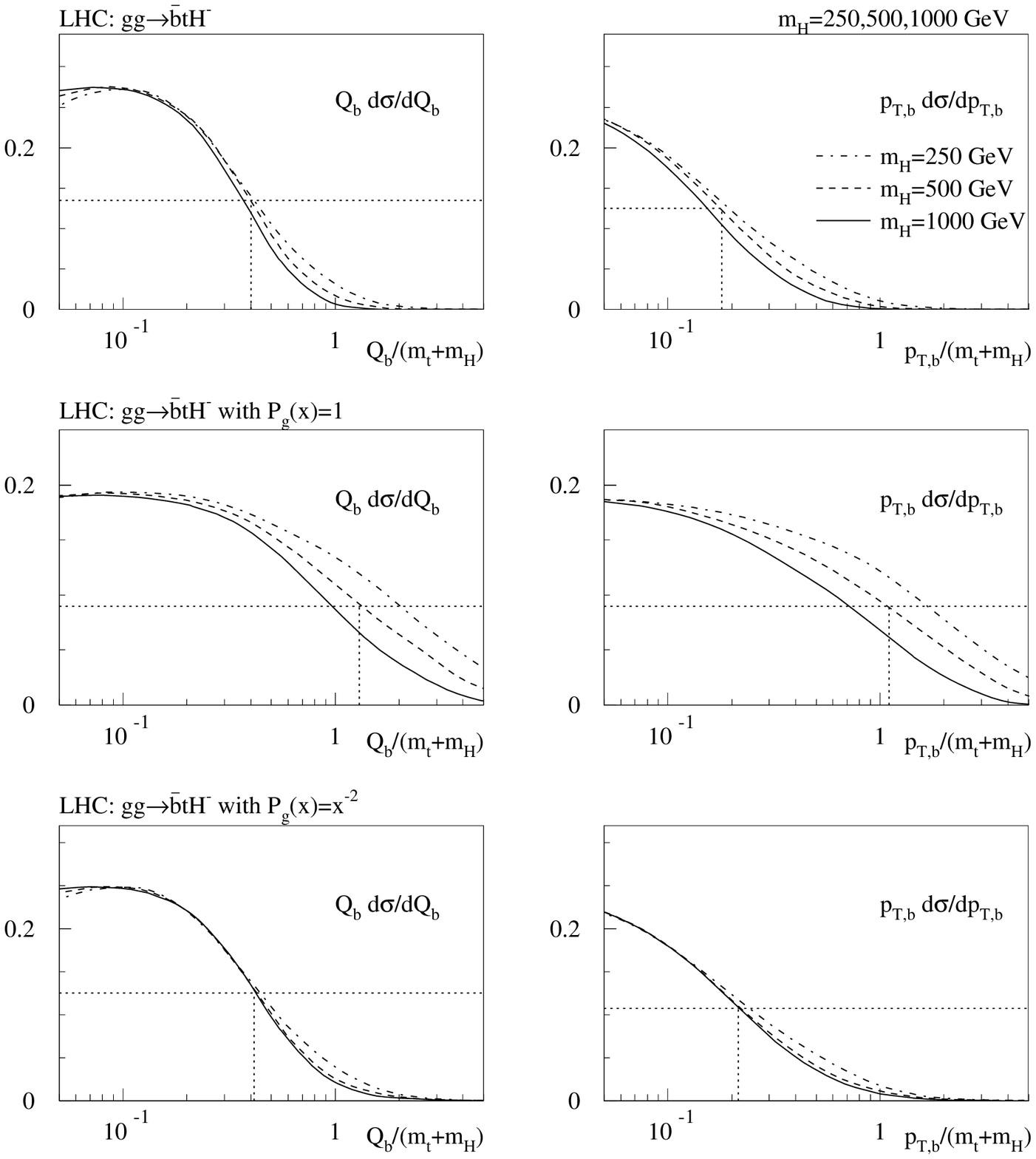}
\end{center}
\vspace*{0mm}
\caption[]{\label{fig:bth} Normalized distributions for the hadronic
charged Higgs boson production process, for the complete gluon density
(first row), for a constant gluon density (second row), and for the
approximate gluon density $P(x)=1/x^2$ (third row). The left column
shows the bottom quark virtuality distribution, the right column the
bottom quark transverse momentum. The normalization for the largest
Higgs boson mass is by the total rate; for all other masses the curves
are normalized such that their maxima coincide. The normalization
factors for the virtuality and the transverse momentum are
identical. We note that for a comparison with the approximate form
$F(\tau)$ we have to identify $M=m_t+m_H$.}
\end{figure}

Let us now turn to a detailed discussion of the function $F(\tau)$,
shown in Fig.~\ref{fig:func}. As mentioned above, 
$F(\tau)$ is a correction to the known asymptotic behavior of the
differential hadronic cross section with respect to the bottom quark
virtuality.  As in Ref.~\cite{mine}, we show the (normalized)
curves for $Q_b d\sigma/dQ_b$, which we numerically obtain for the
exclusive $\bar{b}tH^-$ production process, in Fig.~\ref{fig:bth}. The
first thing we notice from the exact results in Fig.~\ref{fig:bth} is
that, as a function of $Q_b/M$, the curves for different Higgs boson
masses are almost identical. The only major difference arises from the
finite bottom mass effects, since their onset does not scale with
$Q_b/M$, but with $m_b/Q_b$. On the other hand, these mass effects are
understood and of no relevance to our argument, which is concerned
with the {\sl upper} end of the asymptotic behavior in $Q_b$. We
roughly indicate this upper end of the plateau with a dotted line,
where $Q_b d\sigma/dQ_b$ has dropped to half of its plateau value.  In
the first row in Fig.~\ref{fig:bth} we observe how the plateau in the
virtuality does not at all extend to $Q_b=M$. The asymptotic
approximation of the virtuality distribution is valid only up to
values $Q_b^{\rm max} \sim M/2.5$. In the second row we see that this
picture changes when we ignore the gluon densities, but keep
everything else, like in the complete numerical analysis: now the
asymptotic behavior extends to $Q_b \gtrsim M$. In other words, the
short plateau in $Q_b d\sigma/dQ_b$ is an effect of the steep gluon
density in the proton.  In the last row we also show how the
approximation of the gluon luminosity ${\cal L}_{gg}(x_1,x_2) \sim
1/(x_1 x_2)^2$, which is a major ingredient we use to derive the
approximate form $F(\tau)$, works very well for the virtuality
distribution.\smallskip

Since our main interest is the size of the bottom quark virtuality
plateau, we need to compare the approximate form of $F(\tau)$ in
Fig.~\ref{fig:func} with the exact distribution $Q_b d\sigma/dQ_b$ in
the first row of Fig.~\ref{fig:bth}. In all figures we include the
line which indicates where $Q_b d\sigma/dQ_b$ has decreased to half the
plateau value. This is as good a measure for the extension of the
plateau as any other. For the discussion of the approximate function
$F(\tau)$ we prefer a better suited measure: the turning point of
$F(\tau(\log Q_b/M))$
\begin{equation}
\frac{d^2 F(\tau(Q_b^2))}{d (\log Q_b)^2} \Bigg|_{Q_b^{\rm max}} = 0.
\label{eq:deriv}
\end{equation}
In Fig.~\ref{fig:func} and Fig.~\ref{fig:bth} we see that this
definition gives essentially the same $Q_b^{\rm max}$ values as the
dotted line. The numerical values for $Q_b^{\rm max}$ which we compute
from the definition of $F(\tau)$ are given in Tab.~\ref{tab:qmax} for
different values of $j$. The case $j=2$ corresponds to the case with
constant parton densities. It is in very good agreement with what we
see in the second row of Fig.~\ref{fig:bth}. The case $j=4$ should
give the extension of the plateau for $gg$- and $qb$-initiated
processes, for example the charged Higgs boson production
process. We see that the approximation $\overline{|{\cal M}|^2} = S^2
\sigma_0/Q_b^2$ does not give a perfect prediction of $Q_b^{\rm max}$,
as it leads to $Q_b^{\rm max} \sim M/1.8$. For interfering $s$ and $t$
channel diagrams in the production process $gg \to \bar{b} X_M$, the
common denominator in the differential cross section typically becomes
$1/(s Q_b^2)$, while the numerator is dominated by the heavy mass. In
our simple approximation we did not take into account this additional
factor $1/s$, which increases $j$ to 5 and moves the turning point to
$Q_b^{\rm max} \lesssim M/2$ and therefore much closer to the values
seen in Fig.~\ref{fig:bth}.\bigskip

To summarize this section: we have shown that partonic phase space
effects are responsible for the maximum value $Q_b^{\rm max}$, up to
which the asymptotic behavior of the hadronic cross section with
respect to the bottom quark virtuality is valid. Our very simple
approximation agrees with the numbers we obtain for the full hadronic
process $gg \to \bar{b}tH^-$ in Fig.~\ref{fig:bth}:
\begin{equation}
Q_b^{\rm max} = C_Q \, M \qquad \qquad \qquad
C_Q \Big|_{\rm approx} \sim \frac{1}{2} \qquad \qquad
C_Q \Big|_{btH} \sim \frac{1}{2.5}
\label{eq:res1}
\end{equation}

\begin{table}
\begin{center}
\begin{tabular}{rr|rc|rc|rc|}
$\sqrt{S}$ & $M$  & $Q_b^{\rm max}$ & $Q_b^{\rm max}/M \; (j=4)$ 
                  & $Q_b^{\rm max}$ & $Q_b^{\rm max}/M \; (j=5)$ 
                  & $Q_b^{\rm max}$ & $Q_b^{\rm max}/M \; (j=2)$ \\
2000  & 130  &  74    & 1/1.76   &  64    & 1/2.03   & 123    & 1/1.06   \\
      & 250  & 142    & 1/1.76   & 123    & 1/2.03   & 235    & 1/1.06   \\
      & 500  & 282    & 1/1.77   & 245    & 1/2.04   & 463    & 1/1.08   \\
      & 1000 & 559    & 1/1.79   & 488    & 1/2.05   & 908    & 1/1.10   \\
14000 & 130  &  72    & 1/1.81   &  64    & 1/2.03   & 119    & 1/1.09   \\
      & 250  & 138    & 1/1.81   & 121    & 1/2.07   & 221    & 1/1.13   \\
      & 500  & 271    & 1/1.85   & 238    & 1/2.10   & 416    & 1/1.20   
\end{tabular}
\end{center}
\caption[]{\label{tab:qmax} Maximum values for the bottom quark
virtuality at the Tevatron and at the LHC, as defined in
eq.(\ref{eq:cxn3}) and eq.(\ref{eq:deriv}). The values of $j$
correspond to the power of $x$ in eq.(\ref{eq:cxn3}), as it arises
from the $x_1 x_2$ behavior of the partonic cross section.}
\end{table}

\section{Asymptotic Behavior in the Bottom Quark Transverse Momentum} 
\label{sec:plateau_pt}

From the discussion of the bottom quark virtuality one would expect to
be able to follow a similar set of arguments for the bottom quark
transverse momentum. Unfortunately, the phase space parameterization
reflects the fact that the hadronic cross section factorizes in the
virtuality and not in the transverse momentum. Instead, we choose a
different path: we know that the asymptotic behavior $d\sigma/dp_{T,b}
\propto 1/p_{T,b}$ has to hold for small transverse momenta. In that
regime the longitudinal momentum of the outgoing bottom quark in the
center-of-mass system will be much larger than the transverse
momentum: $\rho \equiv p_{z,b}/p_{T,b} \gg 1$. On the other hand, we
want to push this approximation to its limits, $\rho \ll 1$. The
general relation between the virtuality and the transverse momentum of
the bottom jet for the production of a heavy system $X_M$ at threshold
is:
\begin{equation}
\frac{Q_b^2}{M^2} = \frac{p_{T,b} \, \sqrt{s}}{M^2} \, 
                 \left( \sqrt{1+\rho^2} - \rho \right)
                 \sim \frac{p_{T,b}}{M}
\label{eq:qpt1}
\end{equation}
In the intermediate step we have assumed that in the limit of large
bottom quark virtuality and large transverse momentum the bottom jet
has dominantly a transverse momentum direction, \ie that the bottom
jet is central in the detector. We now require that the phase space
region which forms the upper end of the plateau in the virtuality
($Q_b^{\rm max}$) is also responsible for the upper end of the
transverse momentum plateau ($p_{T,b}^{\rm max}$). From the
approximate result in eq.(\ref{eq:res1}) we obtain
\begin{equation}
\mu_{F,b} = p_{T,b}^{\rm max} = C_p \, Q_b^{\rm max}  
          = C_p \, C_Q \, M
                   \qquad \qquad \qquad 
C_p \Big|_{\rm approx} =
C_Q \Big|_{\rm approx} = \frac{1}{2}
\label{eq:res2}
\end{equation}\bigskip 

To understand this effect in more detail, we now keep $\rho$ as a free
parameter and assume that the heavy system $X_M$ be produced at
threshold. We find:
\begin{equation}
p_{T,b} = \frac{s-M^2}{2 \sqrt{s} \sqrt{1+\rho^2}}
\label{eq:ptq2}
\end{equation}
We can solve the equation for $\sqrt{s}$ and insert it into
eq.(\ref{eq:qpt1}), which leaves us with:
\begin{equation}
\frac{Q_b^2}{M^2} = \frac{p_{T,b}^2}{M^2} \; 
                  \left( 1 + \rho^2 - \rho \sqrt{1+\rho^2} \right) \;
                  \left( 1 + \sqrt{1 + \frac{M^2}{p_{T,b}^2 (1+\rho^2)}} \,
                  \right)
\label{eq:qpt}
\end{equation}

This function has two limiting cases: for small transverse momentum
($\rho \gg 1$), $p_{T,b}$ scales with the virtuality $p_{T,b} \sim
Q_b$. This means that the plateau in the transverse momentum will
extend to the same value as the plateau in the virtuality. This
observation suggests that (at least for not too large transverse
momenta) the phase space region which dominates the high end of the
plateau in $Q_b$ will also be responsible for the upper end of the
plateau in $p_{T,b}$. We have confirmed this assumption for charged
Higgs boson production explicitly. In contrast, for very large
transverse momentum ($\rho \ll 1$), a correction to the linear
relation between $p_{T,b}$ and $Q_b$ occurs. This correction pushes
$p_{T,b}$ to smaller values, and the corrections becomes bigger for
small values of $p_{T,b}/M$, which is what we expect from the shifted
and softened transverse momentum plateau in
Fig.~\ref{fig:bth}.\medskip

\begin{figure}[t] 
\begin{center}
\includegraphics[width=7.0cm]{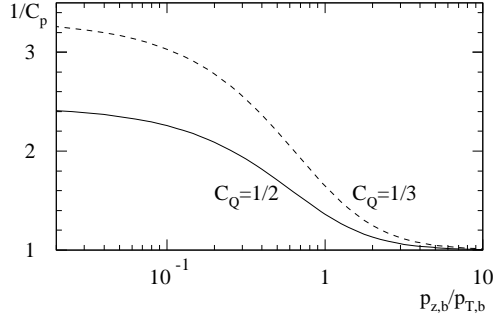}
\end{center}
\vspace*{0mm}
\caption[]{\label{fig:qpt} The numerical solution of eq.(\ref{eq:qpt})
for two different values of $C_Q=Q_b/M$, giving $C_p=p_{T,b}/Q_b$ as a
function of $\rho=p_{z,b}/p_{T,b}$. The longitudinal momentum of the
bottom quarks is defined in the parton center-of-mass system.}
\end{figure}

To translate $Q_b^{\rm max}$ into $p_{T,b}^{\max}$ we evaluate
eq.(\ref{eq:qpt}) numerically. The left hand side we substitute by
$C_Q$, which according to the previous section assumes numerical
values of $1/2$ to $1/3$. Fig.~\ref{fig:qpt} shows the corresponding
values of $C_p = p_{T,b}^{\rm max}/Q_b^{\rm max}$ for different $\rho$
values. Indeed $C_p=1$ holds for small transverse momentum $\rho
\gtrsim 5$. For all other values of $\rho$ the transverse momentum is
always considerably smaller than the virtuality. In other words: if we
want the transverse momentum to be as large as possible {\sl for a
given virtuality}, we would have to make $\rho$ large, \ie make the
longitudinal momentum even larger. The limiting factor will be once
more the steep gluon luminosity. In the opposite regime $\rho \ll 1$,
which will be preferred by the gluon density, we find a very
substantial reduction factor $C_p \lesssim 1/2$.\smallskip

In the right panel of Fig.~\ref{fig:para_bth} we plot the $\rho$
distribution for exclusive charged Higgs boson production. It is
clearly peaked at small values of $\rho$, which means large values of
$p_{T,b}$.  The peak becomes considerably more pronounced if we only
allow virtualities above $M/5$, which limits the phase space to the
transition region in the virtuality plateau. According to
Fig.~\ref{fig:qpt}, the region $\rho \ll 1$ leads to transverse
momenta much smaller than the virtuality.  Even though the actual
distribution in Fig.~\ref{fig:para_bth} peaks at $\rho=0$ these events
will not anymore contribute to the upper end of the plateau in
$p_{T,b}$; instead the plateau in $p_{T,b}$ will be softened. At the
opposite end, $\rho \gtrsim 2$, we would expect a negligible suppression
factor $C_p$ and large transverse momenta from virtualities around
$Q_b^{\rm max}$, but the actual distribution shows how the gluon
luminosities cut heavily into this region. What is left is the
intermediate region $\rho =0.5 \cdots 1.5$, which interpolates between
the extremes and contributes most to the upper end of the plateau in
$p_{T,b} d\sigma/dp_{T,b}$. Even though the numerical details used in
this argument are not process independent, the fact that one has to
interpolate between the two extreme regions in phase space is
completely general.\bigskip

This leaves us with two conclusions concerning the connection between
the asymptotic regions in the bottom quark virtuality and transverse
momentum. First, the plateau will not just translate from $Q_b
d\sigma/dQ_b$ to $p_{T,b} d\sigma/dp_{T,b}$. Instead, it will be
softened. Second, we can extract our approximate prediction for
$C_p$ from Fig.~\ref{fig:qpt} (keeping in mind $C_Q \sim 1/2.5$) and
compare it with what we find in Fig.~\ref{fig:bth}:
\begin{equation}
\mu_{F,b} = p_{T,b}^{\rm max} = C_p \, Q_b^{\rm max}  
                   \qquad \qquad \qquad 
C_p \Big|_{\rm approx} \sim \frac{1}{1.4} \cdots \frac{1}{2} \qquad \qquad
C_p \Big|_{btH} \sim \frac{1}{2} 
\end{equation}\bigskip 

Together with the results presented in Section~\ref{sec:plateau_q} we
now understand that for charged Higgs boson production the plateau in
$p_{T,b} d\sigma/dp_{T,b}$ does not extend to values $p_{T,b}^{\rm
max} \sim M$. Making use essentially of phase space effects we instead
find $p_{T,b}^{\rm max} \sim M/4$. This result confirms the value
$p_{T,b}^{\rm max} \sim M/5$, which we find directly from the
exclusive process $gg \to \bar{b}tH^-$~\cite{mine}, which implies that
using the naive bottom quark factorization scale $\mu_{F,b} = M$ will
overestimate the cross section considerably. On the other hand, higher
order QCD corrections~\cite{charged_bb,zhu,mine,scott2,neut_bb_nnlo}
soften the dependence on the factorization scale considerably.

We can also turn this argument around: the bottom parton approach
means integrating over the additional bottom quark phase space and
resumming the logarithms including the transverse momentum. Going back
to eq.(\ref{eq:excl_asymp}), we see that the terms which we resum are
at maximum $\alpha_s \log [(p_{T,b}^{\rm max}/m_b)^2]$. For a charged
Higgs boson of mass $500\gev$ this gives $6.7 \alpha_s$. For a
threshold mass $M=130\gev$ the logarithm only yields moderate $3.5
\alpha_s$, using $p_{T,b}^{\rm max} = M/5$.

\section{Similar and Not So Similar Processes}
\label{sec:others}

\begin{figure}[t] 
\begin{center}
\includegraphics[width=14.0cm]{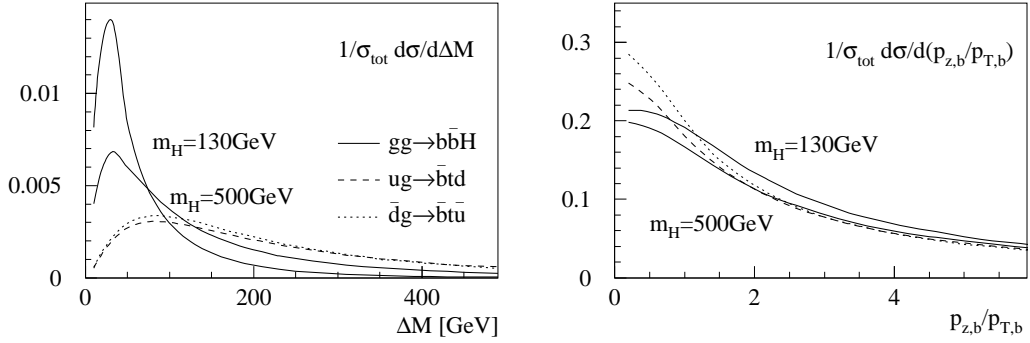}
\end{center}
\vspace*{0mm}
\caption[]{\label{fig:para_others} Normalized distributions for
hadronic neutral Higgs boson production $gg \to \bar{b}b\Phi$ and
single top quark production $qg \to \bar{b}tq'$. Left: difference
between the invariant mass of the heavy system and its threshold
mass. Right: ratio of the longitudinal and transverse momentum of the
bottom jet $\rho = p_{z,b}/p_{T,b}$ in the parton center-of-mass
system.}
\end{figure}

\begin{figure}[t] 
\begin{center}
\includegraphics[width=14.0cm]{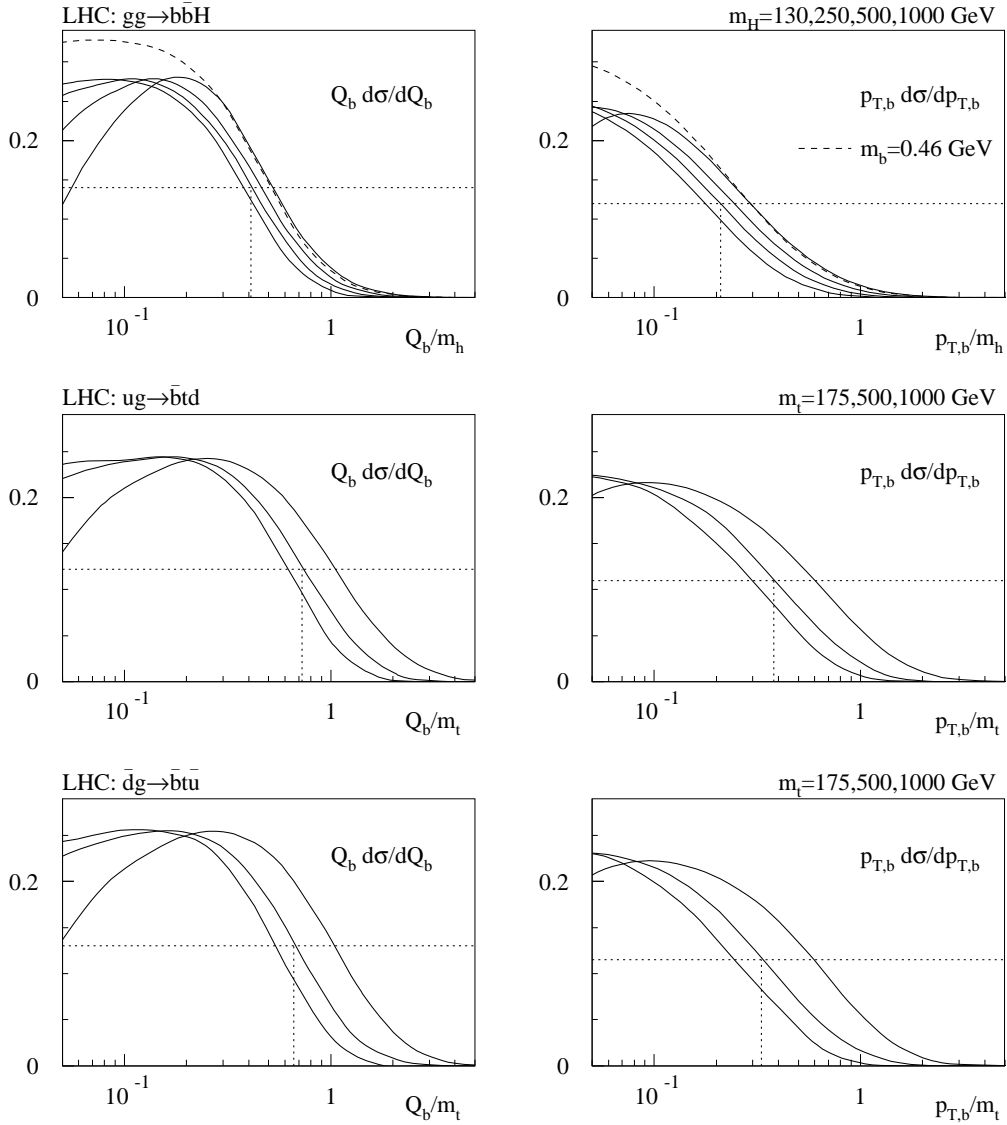}
\end{center}
\vspace*{0mm}
\caption[]{\label{fig:others} Normalized distributions for the
hadronic neutral Higgs boson production and for exclusive single top
quark production at the LHC. The normalization for the largest Higgs
boson mass is by the total rate; for all other masses the curves are
normalized such that the virtuality distributions coincide at their 
maxima. The normalization factors for the virtuality and the
transverse momentum are identical. The dashed curve uses a
mathematical cutoff $0.46\gev$ for the bottom quark mass and $130\gev$
for the Higgs boson mass. It is normalized to match the curve for the
physical bottom quark mass for large virtuality. The curves are
ordered on their downward slopes by decreasing final state mass
towards larger values of $Q_b^{\rm max}$ and $p_{T,b}^{\rm max}$. In the
case of single top quark production we use three different top quark
masses to illustrate the dependence on the final state mass, even
though the top mass is a measured parameter
%%%~\cite{vernon}
.}
\end{figure}

\begin{figure}[t] 
\begin{center}
\includegraphics[width=14.0cm]{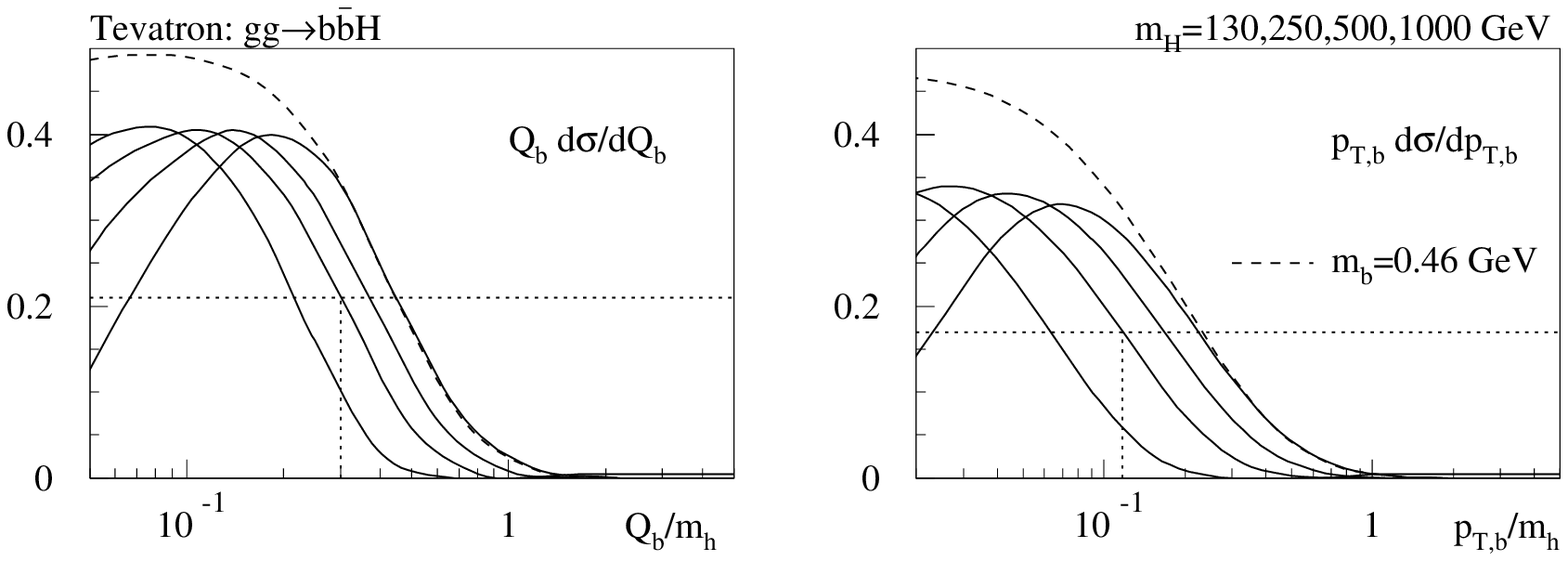}
\end{center}
\vspace*{0mm}
\caption[]{\label{fig:tev} Normalized distributions for the hadronic
neutral Higgs boson production at the Tevatron. The normalization for
the largest Higgs boson mass is by the total rate; for all other
masses the curves are normalized such that the virtuality
distributions coincide at their maxima. The normalization factors for the
virtuality and the transverse momentum are identical. The dashed curve
uses a mathematical cutoff $0.46\gev$ for the bottom quark mass and
$130\gev$ for the Higgs boson mass. It is normalized to match the
curve for the physical bottom quark mass for large virtuality. The
curves are ordered on their downward slopes by decreasing final state
mass towards larger values of $Q_b^{\rm max}$ and $p_{T,b}^{\rm max}$.}
\end{figure}

\subsection{Neutral Higgs Boson Production}

In the previous sections we derived the appropriate bottom parton
factorization scale for the associated production of a charged Higgs
boson and a top quark. For different reasons this process is
particularly well suited for the bottom parton description: there is
only one bottom parton, the top--Higgs system is very heavy, it will
be produced close to threshold, and (for those reasons) it will be
slow-moving and central in the detector. A process which is
particularly important for light as well as for heavy supersymmetric
Higgs bosons is the production of a neutral scalar in association with
two bottom quarks, which at the different level of inclusive versus
exclusive description reads~\cite{neut_bb_excl,neut_bb_nnlo}:
\begin{equation}
gg \to \bar{b} b \Phi \qquad \qquad \qquad
bg \to b \Phi \qquad \qquad \qquad
b\bar{b} \to \Phi
\end{equation}\bigskip

A phenomenological interesting aspect is that these kind of channels
with at least one tagged bottom jet~\cite{scott2,scott1} can prove an
enhanced bottom Yukawa coupling. We know that just like in the charged
Higgs boson case, the factorization scale of the bottom parton has to
be chosen well below $m_\Phi$~\cite{mine,scott2}. The first reason is
that again a heavy system ($X_M=b\Phi$) is produced close to
threshold, Fig.~\ref{fig:para_others}. In Fig.~\ref{fig:others} we
show the plateau in $Q_b d\sigma/dQ_b$, similarly to the figures in
Ref.~\cite{scott2}, and obtain $Q_b^{\rm max} \sim M/2.5$. Like for
the charged Higgs boson case we see that the curves are nearly
degenerate for different Higgs boson masses, \ie $Q_b^{\rm max}
\propto M$.  For the neutral Higgs bosons an additional curve is
included in Fig.~\ref{fig:others}, assuming a $130\gev$ light scalar
MSSM Higgs boson. In this case the plateau is not particularly wide,
since the bottom mass effects bend down the curves at fairly large
values of $Q_b/M$. On the other hand, these effects are understood, so
that the altered shape can be treated just like a plateau. Moreover,
the dominant effects will arise from the upper end of the curve, where
the logarithms are largest.\smallskip

Comparing Fig.~\ref{fig:bth} with Fig.~\ref{fig:others}, we see that
the extension of the asymptotic behavior in the bottom quark
virtuality at the LHC is just the same for charged Higgs boson
production ($bg \to tH^-$) as for neutral Higgs boson production ($bg
\to b\Phi$), for similar heavy state masses. An aspect we did not
discuss in the charged Higgs boson case is that the plateau seems to
extend to slightly larger virtualities for smaller threshold masses,
in particular for a $130\gev$ neutral Higgs boson. The reason is that,
in general, $X_M$ will be produced relatively closer to threshold for
heavier states, \ie $\Delta M/M$ becomes smaller, even though
Fig.~\ref{fig:para_bth} and Fig.~\ref{fig:para_others} show that in
absolute numbers $\Delta M$ becomes slightly larger for heavier states
$X_M$. \bigskip

Up to this point we did not have to specify the collider energy in our
approximation. Moreover, charged Higgs boson production is only
relevant at the LHC. The distributions for neutral Higgs boson
production at the Tevatron are given in Fig.~\ref{fig:tev}. While for
small Higgs boson masses of $130\gev$ the bottom quark virtuality 
plateau looks very similar to corresponding curves for the LHC,
our picture starts to break down for very large Higgs boson masses. In
that case the limited hadronic collider energy does not allow
production of heavy states plus a bottom jet with sizeable
virtuality/transverse momentum.  Of course, the case of a $500\gev$
Higgs boson at the Tevatron is phenomenologically irrelevant. On the
other hand, we learn that implicitly we have used another
approximation in our discussion of the LHC processes: that the
production rate even for large virtuality is limited only by the
parton densities and the parton energy, never by the hadronic collider
energy. Implicitly we check this requirement in Fig.~\ref{fig:bth},
when we confirm $Q_b^{\rm max} \sim M$ after neglecting the effect of
parton densities.\bigskip

Last but not least, the question is what will happen for the
completely inclusive Higgs boson production ($b\bar{b} \to
\Phi$)? From Ref.~\cite{scott2} we know that the argument for the
plateau in $Q_b d\sigma/dQ_b$ works just the same way as before,
except that we now use the semi-exclusive process $bg \to b\Phi$ to
compute $Q_b^{\rm max}$. Using the approximation described in
Section~\ref{sec:plateau_q} we understand what happens: the mass of
the heavy system is still $M = m_\Phi$, but the gluon which splits
into a bottom quark pair now sees a bottom parton on the other side
instead of the gluon in the completely exclusive process. In
Fig.~\ref{fig:lumis} we see that the approximations ${\cal L}_{bg}
\sim 1/(x_1 x_2)^2$ and $x_1 \sim x_2$ still work fine. In a way this
is a consistency check, of course. A significant change in the
behavior of the bottom partons between the first bottom jet and the
second bottom jet to be integrated over in the process $gg \to
b\bar{b}\Phi$ would pose a serious problem for factorization in
general. This result is in very good agreement with the results of the
NNLO QCD calculation for this process: the small bottom quark
factorization scale indeed yields perturbative corrections which are
well under control~\cite{neut_bb_nnlo}.\medskip

The second step is again the transition from the plateau in
virtuality to that in transverse momentum. Not surprisingly, from the
exclusive process we numerically find the same behavior as for the
charged Higgs boson, Fig.~\ref{fig:bth}. Our argument in
Section~\ref{sec:plateau_pt} also works just the same way for neutral
Higgs bosons. From the exclusive process we numerically find
$\mu_{F,b} \sim M/5$, which we can understand in complete analogy to
the charged Higgs boson case in the previous sections.

\subsection{Single Top Quark Production}

In contrast to the Higgs boson production mechanisms, the bottom
parton picture in single top quark
production~\cite{single_top,single_top_nlo}
\begin{equation}
ug \to \bar{b}td             \qquad (ub \to td) \qquad \qquad \qquad 
\bar{d}g \to \bar{b}t\bar{u} \qquad (\bar{d}b \to t \bar{u})
\end{equation}
has never posed a problem.  The difference between the two above
processes is that the first one will involve valence quarks at the
LHC, while the second one will not. Looking back, Fig.~\ref{fig:lumis}
shows that these channels should look slightly different, if our
argument in Section~\ref{sec:plateau_q} is correct.\smallskip

In the left column in Fig~\ref{fig:others} we see how the single top
quark case differs from the Higgs boson production. First of all, the
plateau in the virtuality extends considerably further, typically to
$Q_b^{\rm max} \gtrsim M/1.5$. This is in agreement with the less
steep parton densities for the quarks, which the splitting gluon
sees. If we have a closer look, we even see that the plateau extends
further in the case of an incoming valence quark than for a sea quark,
which is in agreement with the approximate parton densities,
Fig.~\ref{fig:lumis}.

However, our approximation has to be looked at with some care, since
now we cannot assume $x_1 \sim x_2$ anymore. Furthermore, in
Fig.~\ref{fig:para_others} we see that the single top quark production
does not at all happen close to threshold. It peaks around $\Delta M
\sim m_W$, which reflects the fact that one could integrate over the
phase space of the outgoing jet and regard the single top quark
process as $bW$ scattering. This gives the outgoing jet a transverse
momentum kick of the order of the $W$ boson mass. The fact that the
invariant mass of the heavy system $X_M = tj$, as it appears in
eq.(\ref{eq:cxn2}), is easily twice the threshold mass, again
contributes to the larger values of $Q_b^{\rm max}$. Finally, as
discussed in the context of neutral Higgs bosons, production away
from threshold lifts the degeneracy of different values of $Q_b^{\rm
max}/M$, pushing the lighter states to higher values of $Q_b^{\rm
max}$.\smallskip

We will not discuss the transition from the bottom quark
virtuality to the transverse momentum in any detail. From
Fig.~\ref{fig:others} we see that the plateau is softened and
$p_{T,b}^{\rm max} \sim Q_b^{\rm max}/2$, as in all other processes
discussed before: the interpolation argument presented in
Section~\ref{sec:plateau_pt} describes the single top quark production
perfectly well. This part of our argument indeed holds independently
for all processes considered in this paper.\bigskip

Recently, a similar issue of bottom partons was
discussed~\cite{kuehn} in the framework of single top quark production
at a linear collider $e\gamma \to \bar{\nu} t \bar{b}$. The authors
find sizeable differences between the finite $m_b$ prediction and the
(massless) structure function approach, predominantly close to
threshold $\sqrt{s} \lesssim m_t+m_b+10\gev$. These differences can in
part be traced back to phase space effects. From
Fig.~\ref{fig:para_bth} and Fig.~\ref{fig:para_others} we see that
this region of phase space contributes little to the Higgs boson
sample at the LHC, after we convolute the partonic cross section with
the gluon densities, integrating over the entire partonic energy
range. It will have even less impact on the total rate once a minimum
transverse momentum of the Higgs boson decay products is
required. While for a linear collider the bottom quark mass is an
important source of theoretical uncertainty and the collinear
logarithms (multiplied with $\alpha \sim 1/137$) are under control,
the dominant problem at hadron colliders is the size of the logarithms
(multiplied with $\alpha_s$), which we link to the transverse momentum
spectrum in the exclusive processes.

\section{Conclusions}

Starting from charged Higgs boson production in association with a
top quark, we have investigated processes which can be described using
bottom partons. From the kinematics of the exclusive processes, we
numerically find that the factorization scale of the bottom parton has
to be smaller than the threshold mass or the hard scale in the
process: $\mu_{F,b} \sim M/5$. In two steps we first investigate the
validity of the asymptotic approximation in the bottom quark
virtuality and then in the transverse momentum. The upper limit
$p_{T,b}^{\rm max}$, for which the exclusive cross section is
dominated by collinear bottom quarks and large logarithms $\log
(p_{T,b}^2/m_b^2)$, defines the appropriate value for the factorization
scale of the bottom parton in the inclusive process. We derive the
observed dramatic decrease in the factorization scale as compared to
the hard scale $M$ in a process-independent approach, using only
properties of the phase space and of the parton densities. In this
simple picture we indeed find $\mu_{F,b} \sim M/4$.

This observation resolves the puzzle of the discrepancy between
inclusive and exclusive rates, as it has been present in the
literature. Using an appropriate scale, the difference for example for
the process $b\bar{b} \to \Phi$~\cite{scott2} is not huge and well
understood. Moreover, higher order calculations have been pointing to
small bottom factorization
scales~\cite{charged_bb,zhu,mine,scott2,neut_bb_nnlo}; we understand
how this is caused by the partonic phase space without relying on
numerical arguments based on higher order corrections.\smallskip

Turning around the argument, we can specify how large the logarithms
actually are, which are resummed in the bottom parton picture. Again,
they are smaller than the naive guess $\log (M^2/m_b^2)$ would
indicate. In particular, for a light neutral Higgs boson one can debate
using the (resummed) bottom parton cross section, or just
integrating over the exclusive cross section~\cite{tth_nlo}. For heavier
neutral or charged Higgs bosons the logarithms are certainly large
enough to require a resummation.

%%%%%%%%%%%%%%%%%%%%  ACKNOWLEDGMENTS  %%%%%%%%%%%%%%%%%%%%

\acknowledgements

We would like to thank Michael Kr\"amer for fruitful
discussions. Moreover, we would like to thank Thomas Gehrmann, Dave
Rainwater and Robert Harlander for very helpful comments and the
careful reading of the manuscript. T.P. would like to thank the
Physics Department of the University of Edinburgh, where part of this
work was completed, for their hospitality.  E.B. was partly supported
by the INTAS 00-0679, CERN-INTAS 99-377, and Universities of Russia
UR.02.03.002 grants

%%%%%%%%%%%%%%%%%%%%%%%  REFERENCES  %%%%%%%%%%%%%%%%%%%%%%%

\bibliographystyle{plain}

\end{document}